# Microstructure of Glidcop® AL-60

Yan Xin, Jun Lu, and Ke Han

*Abstract*— Glidcop® is an oxide-particle-dispersion strengthened copper composite that has a combination of high mechanical strength and high electrical conductivity. It has been used as a conductor for 100 T ultrahigh field pulsed magnets by the National High Magnetic Field Laboratory, USA. In the quest for even higher field pulsed magnets, material development is crucial. Since the mechanical properties of a material are often determined by its microstructure, full characterization of the microstructure of Glidcop® is necessary. In this work, we studied the microstructure of Glidcop® AL-60 using both transmission electron microscopy (TEM) and scanning transmission electron microscopy (STEM). We identified both $\alpha$-$Al_2O_3$ and cubic $\eta$-$Al_2O_3$ nanoparticles in AL-60 and investigated their size and density distribution. The small alumina particles $\eta$-$Al_2O_3$ nanoparticles with typical size of 5 to 30 nm are of triangular shape. They have had well defined crystal orientation relationship with the Cu matrix. We observed dislocations pinned by the alumina nanoparticles in cold-drawn wires. We believed that dislocation bypassing alumina particles via Orowan looping was the main strengthening mechanism. We observed microcracks near large particles, demonstrating the detrimental effect of large particles in AL-60.

## I. Introduction

OXIDE dispersion strengthened (ODS) copper has good properties combining high mechanical strength and high electrical conductivity [1]. The ODS copper is commercially available [2] with a trade name of Glidcop® in three grades, AL-15, AL-25, and AL-60. Among them, AL-60 has the highest oxides content of 1.1 wt% and highest strength. Glidcop® is manufactured by a powder metallurgy process, in which powder of Cu-Al alloy and oxidant powder are mixed and heat treated, producing dispersed small aluminum oxide particles inside the Cu matrix by internal oxidation [3]. Glidcop® is resistant to thermal softening and creep, and has high strength high electrical and thermal conductivity. Their main commercial applications are MIG welding tips and resistive welding electrodes. They have special applications in electromagnet systems as conductor material for which high strength is essential to withstand large electromagnetic forces. For example, wires made of AL-60 are used as conductors in the 100 T and the 60 T pulsed magnet at the National High Magnetic Field Laboratory, USA [4], [5]. Other potential applications include nuclear fusion experimental reactors [6] and high strength $Nb_3Sn$ superconducting wire [7].

Mechanical properties of a material, such as its strength, fatigue properties, and fracture properties, are often dictated by its microstructures. Therefore, in order to understand its mechanical and electrical behavior under application condition and to prevent unexpected failure, it is necessary to study the microstructure of Glidcop®, which includes the size and the distribution of the strengthening oxides particles in the copper matrix, as well as the interaction between the particles and the dislocations in the Cu matrix. There are several publications that involves microanalysis of Glidcop® [9],[15],[24] and ODS copper in general [9]-[24] in the literature. Nevertheless, a comprehensive study of the basic microstructures of Glidcop® AL-60 and their correlations with the mechanical properties is still strongly desirable for the materials scientists and mechanical engineers working with this material.

In this paper, we use transmission electron microscope (TEM) and scanning transmission electron microscope (STEM) to study the microstructure of Glidcop® AL-60. The size and phases of aluminum oxide particles are determined. Dislocation loops and microcracks around particles were observed and their implication on mechanical properties is discussed.

## II. Experimental Methods

The material studied in this paper was hot extruded precursor rod of Glidcop® AL-60 made by North American Höganäs. The precursor rod was about 2 cm in diameter with a 1 mm thick pure copper cladding layer. Some precursor rods were cold drawn to a final cross-section of 2.5 mm x 4.8 mm.

Alumina particles were extracted from Glidcop Cu matrix by dissolving Cu in $HNO_3$: $H_2O$ = 1:1 in a beaker for about 12 hours. After Cu was dissolved, the $Al_2O_3$ particles were settled to the bottom of the beaker. By carefully diluting and pouring out the acid solution, alumina particles were obtained with small amount of copper nitrate indicated by the light blue color.

A water suspension of the extracted alumina particles was dried on a glass slide and used as a sample for x-ray diffraction (XRD) analysis by the Rigaku SmartLab II machine with Cu K$\alpha$ radiation.

TEM samples of the extracted alumina particles were made by casting water suspension onto carbon coated 200 mesh TEM sample-grid by using a pipette and let it dry. To make TEM samples from bulk AL-60, focused ion beam (FIB) was used to cut a lamella in Thermal Fisher Scientific Helios G4 Dual-

This work was performed at the National High Magnetic Field Laboratory, which is supported by National Science Foundation Cooperative Agreement *No. DMR-1644779* and the State of Florida. Y. Xin, J. Lu, and K. Han are with the Magnet Science and Technology Division, National High Magnetic Field Laboratory, Tallahassee, FL 32310, USA. Corresponding author (junlu@magnet.fsu.edu)

Digital Object Identifier will be inserted here upon acceptance.

Beam scanning electron microscope (SEM). TEM was performed on a probe-aberration-corrected cold emission JEOL JEM-ARM200cF at 200 kV with point resolution of 0.078 nm, equipped with Oxford Aztec SDD EDS detector. The microstructures of the samples were imaged by high-angular-dark-field scanning transmission electron microscopy (HAADF-STEM) and annular-bright-field STEM (ABF-STEM). Electron energy loss spectroscopy (EELS) was used to determine particle or lamella thickness by using the log-ratio method [25].

## III. RESULTS AND DISCUSSION

### A. Alumina particle shape, size, and distribution

X-ray diffraction pattern of the extracted aluminum oxide powder is shown in Fig.1(a). Two phases of aluminum oxides were identified, namely the α-$Al_2O_3$ (corundum) and the metastable cubic phase η-$Al_2O_3$ (space group 227, lattice constant 0.794 nm). The peaks at low 2θ angles were assigned to a copper nitrate phase which corresponds to the residual of the particle extraction process. The peaks of α-$Al_2O_3$ have peak width (FWHM) in the order of 0.1° which are considerably sharper than those of η-$Al_2O_3$ with FWHM in the order of 1°, indicating significantly larger particle size of α-$Al_2O_3$. Both α-$Al_2O_3$ and η-$Al_2O_3$ have been observed in internal oxidation ODS copper [9], [12], [13], [16], [17], [20]. In principle, the particle sizes can be analyzed by XRD peak width by using Scherrer equation. In practice, however, careful determination of instrument broadening is necessary. Since electron microscopy has the advantage of analyzing the shape and chemical composition as well as the size of these particles, detailed particle characterization was performed by TEM.

TEM-EDS spectra of the extracted particles (not shown here) indicate predominant Al and O peaks with almost no other peaks, confirming that these particles are aluminum oxides. Fig. 1(b) exhibits TEM bright field image of extracted alumina particles. Most of the particles in this figure have a triangular shape with their size ranging from 5 – 30 nm. The corresponding selected area (SAD) electron diffraction pattern was indexed as cubic η-$Al_2O_3$ phase (Fig. 1(c)), consistent with our XRD analysis of the extracted powders. Particles of greater than 200 nm are also present in this sample (Fig. 1(d)). For these large particles, single crystal SAD pattern was obtained and indexed to be [0001] α-$Al_2O_3$ (Fig. 1(e)). Most of the particles measured by us (>500 particles) have sizes between 5 and 30 nm. There are larger particles ranging from 100 to 500 nm. Particles larger than 1 μm was previously observed in AL-60 by SEM [22].

Above analyses clearly indicates, that there are two phases of alumina particles in AL-60, namely small triangular-shaped η-$Al_2O_3$ particles in the size range of 5 to 30 nm, and large irregular shaped α-$Al_2O_3$ particles in the order of 200 nm. This is consistent with what was observed by others [9].

Study of alumina particles that were extracted from AL-60 has the advantage of simple preparation of XRD and TEM samples that is good for statistical study of the particles. However,

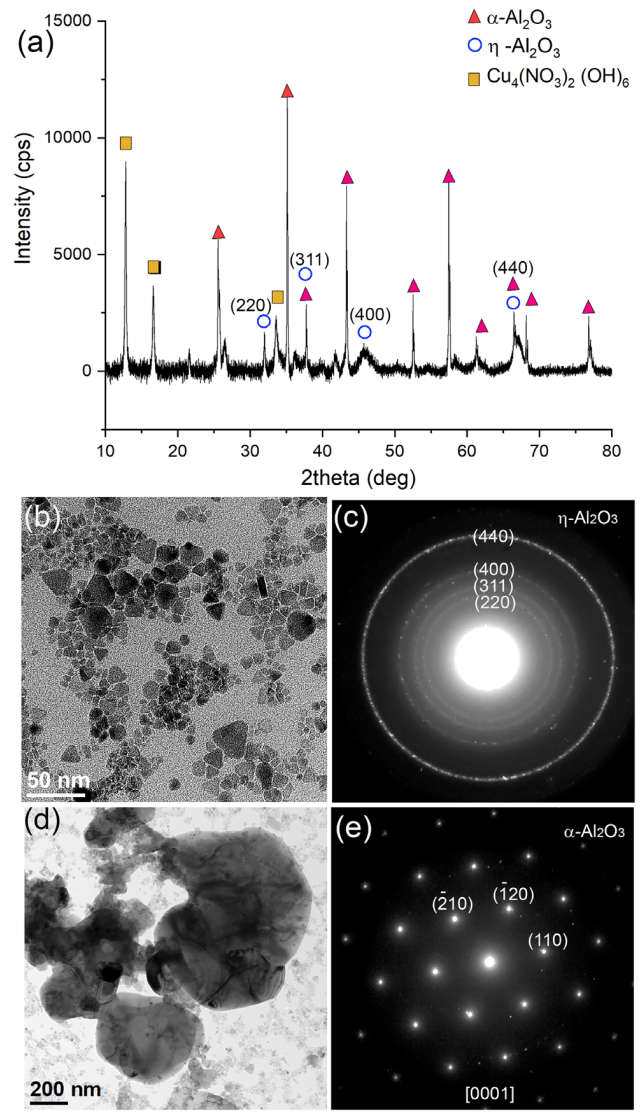

Fig.1 (a) X-ray diffraction pattern of extracted alumina particles. α-$Al_2O_3$ and η-$Al_2O_3$ are present. Only diffractions from η-$Al_2O_3$ are indexed (b) TEM BF image of extracted small alumina particles; (c) Selected area diffraction pattern of (b); (d) TEM BF image of large alumina particles; (e) Selected area diffraction pattern of the largest particle in (d).

it cannot reveal the alumina particle distribution in Cu matrix and the relationship between particles and the Cu matrix. In addition, in the process of removing acidic solution after etching, it is possible that some very fine particles in solution were removed as well. Therefore, it is necessary to make TEM samples out of bulk AL-60. So a few TEM samples were made from AL-60 precursor rod or cold-drawn wire by FIB. Fig. 2(a) is an SEM image of longitudinal cross-section prepared by FIB before this sample was made to a TEM sample. The white speckles are alumina particles. The alumina particle size and density are not uniform. Large particles (50 - 200 nm) have irregular shapes and seem to form a band along the longitudinal direction. Near the large particles, the density of fine particles is very low. Medium sized particles (10 - 30 nm) can be seen throughout the figure. Apparently, the particles do not have a uniform size and are not uniformly distributed. Similar sized particles form bands along the extrusion direction, and the average particle size varies across the horizontal direction of the image.

The non-uniform particle distribution is further demonstrated in HAADF-STEM images Fig. 2(b) and 2(c). HAADF-STEM imaging is a better way to reveal alumina particles in Cu matrix than conventional BF TEM imaging, where the weak particle contrast is usually obscured by diffraction contour bands. Since HAADF image intensity is proportional to the square of atomic number $Z^2$ [26], Al and O with lower Z in alumina particles show dark contrast compared to Cu in the matrix. The regions lack of particles is evident in these two images. This might be the result of incomplete internal oxidation. One of the reasons might be non-uniform mixing of oxidant and Cu-Al powders in AL-60 fabrication process. This suggests that thorough mixing of the ingredient powders in the process is critical to obtain uniformly distributed small alumina particles in AL-60.

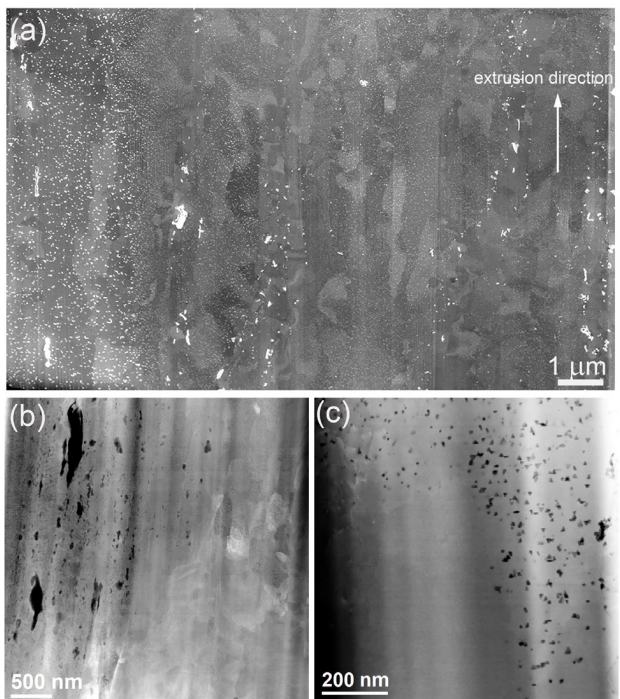

Fig. 2 (a) SEM image of the longitudinal cross-section view of the hot extruded Al-60 wire. White dots are alumina particles. (b) HAAADF-STEM image of a region from a cold drawn wire showing large particles of irregular shapes and region with no alumina particles. (c) Another region from a cold drawn wire with small particles adjacent to no particle region.

Next, we investigated the crystal relationship between the particles and the Cu matrix. Fig. 3(a) is an HAADF-STEM image of small alumina particles in a single copper grain. The electron beam is along Cu [110] direction. The particles in this figure have average size of about 10 nm. Most of these small particles have a triangle shape with two straight edges parallel to Cu {111} planes. The rectangular shaped particles indicated by the arrows, are most likely to be triangular particles lying sideways in the Cu {111} planes that are parallel to the electron beam direction of Cu [110]. These triangular particles seem to be similar to the platelets in AL-60 observed by Ernst et al [9], where the platelet facets are bounded by Cu {111} and lies on

Cu {111} planes with edges of the triangle along <110> directions. Our electron diffraction pattern obtained from the area

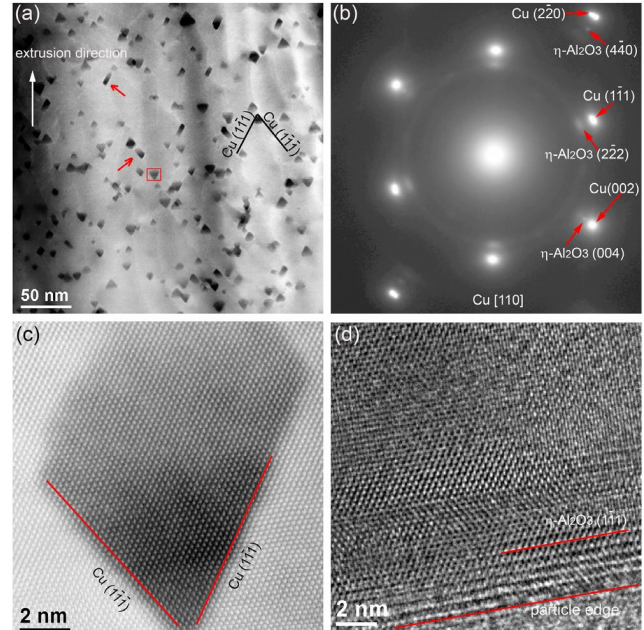

Fig.3 (a) HAADF-STEM image of the small alumina particles in a single Cu grain. The electron beam is along Cu [110] direction. The particles in the red square are examples of triangular-shaped alumina, while the arrows pointed are the particles orientated sideways. (b) Selected area diffraction pattern from area with triangle shaped particles. The weak ring and some diffraction arc are from $Cu_2O$ due to Cu surface oxidation during sample preparation. (c) Atomic resolution HAADF-STEM image of a particle in copper matrix viewed from Cu [110] zone axis. (d) High resolution TEM image of the straight edge of an extracted loose triangle particle.

that includes multiple particles shows that $\eta\text{-}Al_2O_3$ particles have single crystal pattern of [110], which is parallel to [110] of copper matrix (Fig. 3 (b)). In the atomic resolution HAADF-STEM image taken from Cu [110] zone axis (Fig. 3(c)), the atomic columns shown are from Cu, the dark region is alumina particle. Two straight edges of the particle are parallel to Cu {111}. The edge facets of the particle are confirmed to be $\eta\text{-}Al_2O_3$ {111} from the high-resolution TEM image of an extracted triangular shaped particle (Fig. 3(d)). We have previously determined from SAD of the powder TEM sample that these small triangle shaped particles are crystalline $\eta\text{-}Al_2O_3$. It seems that the particle and Cu matrix have a simple cube-on-cube crystal orientation relationship, i.e., $\eta\text{-}Al_2O_3$ {111}// Cu {111}, $\eta\text{-}Al_2O_3$ {001}// Cu {001}, where two copper unit cell (a = 0.362 nm) coincide with one $\gamma\text{-}Al_2O_3$ cell (a = 0.794 nm). The misfit between Cu and $\eta\text{-}Al_2O_3$ particle is about 10% with Cu in tensile stress at the interface. Using EELS log-ratio method, the thickness of the triangular platelet to the length of its side is estimated to be between 1:3 and 1:1. This is consistent with the aspect ratio of the platelets lying on their sides as indicated with arrows in Fig. 3(a).

B. *Dislocations in cold drawn wire*

The high strength of AL-60 is mostly due to the abilities of the particles to block the movement of dislocations. We therefore studied dislocation in cold-drawn wire. In this sample, due to large strain and high dislocation density, the dislocation contrasts in TEM BF images are obscured by bend contour contrast. Whereas annular bright field (ABF) STEM imaging has much less bend contour contrast, so the dislocations can be seen clearly [27]. Meanwhile. HAADF-STEM is a good imaging method for a system that consist of precipitates that have significantly different atomic number Z from their matrix, as in this case. A combination of ABF-STEM and HAADF-STEM images of the same area clearly resolved both dislocations and alumina particles (Figs. 4 (a); (b)). The alumina particles, which corresponds to the dark contrast in Fig. 4(b), and their interactions with the dislocations are revealed. In close-up images (Fig. 4(c)-(e)), dislocations near an alumina particle are shown.

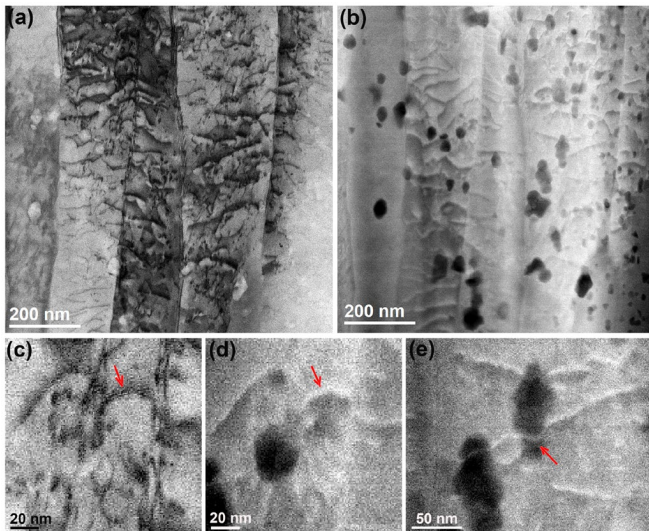

Fig.4 A sample from a cold-drawn AL-60 wire. (a) ABF-STEM image. Dark lines are dislocations. Alumina particles are of light contrast. (b) The HAADF-STEM image of the same region as (a). Alumina particles are of dark contrast. (c) and (d) a pair of ABF and HAADF images showing a half dislocation loop around a particle. (e) HAADF-STEM image showing bowing of a dislocation around an alumina particle.

Small dislocation loops indicated by the arrows circle around the alumina particle, suggesting Orowan strengthening mechanism.

### C. Micro-crack in fractured cold drawn wire

A common fatigue failure mode of a material is through micro-crack initiation and propagation. We made a TEM lamella from a fractured tensile test sample near the fractured surface and found direct evidence of these micro-cracks near the particle/matrix interface. Shown in Fig. 5(a) is an ADF-STEM image of what is left of a large alumina particle, which is about 320 nm wide and 1000 nm long. Micro-cracks, indicated by red arrows, can be clearly seen at both end, which is along the drawing direction. Similar microcracks associated with large particles in AL-60 have been previously observed by SEM [22]. Fig. 5(b) and (c) are EDS elemental maps of the region, which interestingly reveal that only part of the alumina particle is left after FIB cutting. The rest of area of the particle seems to be pure copper which, however, has very fine grain (ranges from 10 to 20 nm) structure definitively different from that of the copper matrix (Fig. 5(a) inset). The origin of this cluster of fine Cu nanoparticle is not clear, but they are usually associated with large alumina particles. This image shows that micro-cracks preferentially started at the large alumina particle/small Cu cluster region. It is conceivable that during the wire drawing with large reduction area, similar micro-cracks may form and propagation resulting macroscopic internal cracks (chevron cracks) or even wire breakage.

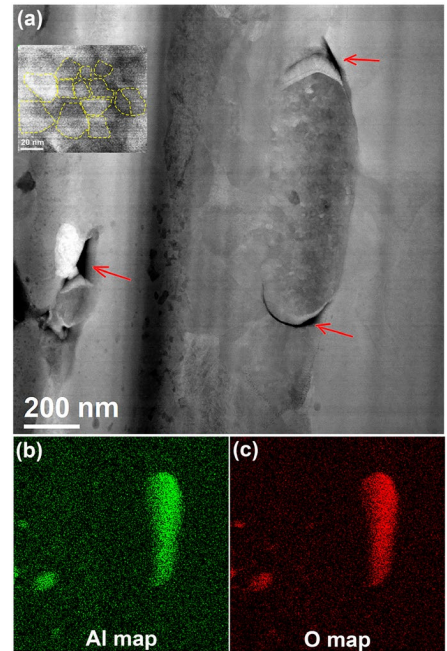

Fig. 5 Sample made near surface of the fracture after a tensile test wire. (a) ADF-STEM image of micro-cracks associated with large alumina particles. Inset: enlarged image from the particle with small grains delineated with dash lines. (b) and (c) EDS elemental mapping of the same region showing the existence of large alumina particles.

### IV. SUMMARY

We performed microstructure analysis on AL-60 samples using x-ray diffraction, transmission electron microscopy (TEM) and scanning transmission electron microscopy (STEM). Alumina particles averaging 200 nm and 10 nm size are identified to be $\alpha$-$Al_2O_3$ and metastable cubic $\eta$-$Al_2O_3$ respectively. The $\eta$-$Al_2O_3$ particles have triangular shape whose crystal lattice has a simple crystal relationship with Cu matrix $\eta$-$Al_2O_3$ {111}// Cu {111}, $\eta$-$Al_2O_3$ {001}// Cu {001}. Particles of different sizes are typically grouped in bands elongated in the extrusion direction of the AL-60 rod. Dislocations loops bent around the alumina nanoparticles are observed in a cold-drawn wire sample suggesting Orowan strengthening mechanism. In a sample made near a fracture surface, we found a large particle with microcracks at its interface with copper matrix. This demonstrates the detrimental effect of large particles in the material.

### ACKNOWLEDGMENT

We thank Vince Toplosky and Robert Goddard for assistance in sample preparations. We acknowledge professors Fumitake